\documentclass[]{ifacconf}

\counterwithin*{section}{part}

\usepackage[square]{natbib} 
\renewcommand*{\cite}[2][]{\citep[#1]{#2}}

\usepackage[utf8]{inputenc}
\usepackage[T1]{fontenc}
\usepackage[english]{babel}

\usepackage{lmodern}

\usepackage{mathtools}
\usepackage{amssymb}
\usepackage{graphicx}
\usepackage[dvipsnames]{xcolor}

\usepackage{physics}
\usepackage{siunitx}

\usepackage{psfrag}

\newcommand*\VGS{V_{\mathrm{GS}}}
\newcommand*\VDS{V_{\mathrm{DS}}}

\newcommand*\Vth{V_{\mathrm{th}}}
\newcommand*\ID{I_{\mathrm{D}}}

\newcommand*\Vsat{V_{\mathrm{sat}}}

\usepackage{pgf,tikz,pgfplots}
\pgfplotsset{compat=1.10}
\usetikzlibrary{shapes,arrows,calc,arrows.meta,
patterns,backgrounds,positioning,fit}
\tikzset{every picture/.style=black}
\usepackage{circuitikz}

\usepackage[nameinlink,capitalise,noabbrev]{cleveref}
\crefformat{equation}{(#2#1#3)}
\Crefformat{equation}{Equation (#2#1#3)}
\crefrangeformat{equation}{(#3#1#4) to~(#5#2#6)}
\crefmultiformat{equation}{(#2#1#3)}%
{ and~(#2#1#3)}{, (#2#1#3)}{ and~(#2#1#3)}

\definecolor{l}{rgb}{0 0 1} 

\begin{document}
\begin{frontmatter}

\title{Bounding Variance and Skewness of Fluctuations in Nonlinear Dynamical Systems with Stochastic Thermodynamics\thanksref{footnoteinfo}}

\thanks[footnoteinfo]{This work was supported by the Research Project "Thermodynamics of Circuits for Computation" of the National Fund for Scientific Research of Belgium (FNRS) and of Luxembourg (FNR).}

\author[]{Jean-Charles Delvenne and Léopold Van Brandt} 

\address[]{The authors are with the Institute for Information and Communication Technologies, Electronics and Applied Mathematics (ICTEAM) at Université catholique de Louvain, Belgium.
\newline
email: jean-charles.delvenne@uclouvain.be
}

\graphicspath{{figures/}}

\begin{abstract}     
           
Fluctuations arising in nonlinear dissipative systems (diode, transistors, chemical reaction, etc.) subject to an external drive (voltage, chemical potential, etc.) are well known to elude any simple characterisation such as the fluctuation-dissipation theorem (also called Johnson-Nyquist law, or Einstein's law in specific contexts). Using results from stochastic thermodynamics, we show that the variance of these fluctuations exceeds the variance predicted by a suitably extended version of Johnson-Nyquist's formula, by an amount that is controlled by the skewness (third moment) of the fluctuations. As a consequence, symmetric fluctuations necessarily obey the extended Johnson-Nyquist formula. This shows the physical inconsistency of Gaussian approximation for the noise arising in some nonlinear models, such as MOS transistors or chemical reactions. More generally, this suggests the need for a stochastic nonlinear systems theory that is compatible with the teachings of thermodynamics. 
\end{abstract}


\end{frontmatter}

\section{Introduction}
\label{sec:introduction}
Systems in physics and engineering are often composed as the interconnection of conservative elements (capacitances, springs, etc.) with purely dissipative elements (resistances, friction, diffusion, etc.). The dissipative elements are seen as ideal energy reservoirs. Although at macroscopic scale these reservoirs are seen as always absorbing energy, due to microscopic reversibility they are allowed to also release energy to the system, in the form of a random white noise signal. This randomness, which cannot be neglected at the mesoscopic scale (typically at the nanometer scale), obeys thermodynamic laws that are well understood in the linear regime around equilibrium: Einstein's law for diffusion \cite{einstein1905motion}, Johnson-Nyquist's law for electric resistances \cite{Johnson1928,Nyquist1928}, and more generally the fluctuation-dissipation theorem \cite{kubo1966fluctuation} make a link between the mean response (first moment) and the (co)variance of the response (second moment) to an external excitation.

Random fluctuations in nonlinear devices are  more elusive. It has been pointed out in various contexts that no universal simple relation between mean and variance for nonlinear devices  can be found \cite{gupta1982thermal,van1963thermal,van1973nyquist}.

Stochastic thermodynamics has emerged in the late 1990s as the rewriting of the laws of thermodynamics, 
taking into account the random fluctuations of the thermodynamic variables, including their higher-order moments and not just the mean \cite{van2013stochastic}.
A flagship result is the fluctuation relation \cite{jarzynski1997nonequilibrium, crooks1999entropy}, which can be seen as a generalisation of the fluctuation-dissipation valid in broad range of circumstances even far from equilibrium. It involves all moments of fluctuations. The  consequences of the fluctuation relation and other sibling results, both for the general theory and applications in various fields are far-reaching and still unfold today. For instance the recent Thermodynamic Uncertainty Relations (TUR) offer general lower bounds on the variance of observables in terms of the entropy production
\cite{barato2015thermodynamic,horowitz2020thermodynamic,falasco2020unifying}.

The skewness, the third moment measuring the asymmetry of fluctuations around the mean, has been been given comparatively little attention \cite{Salazar2022,Wampler2021}.


In this paper we use these results from stochastic thermodynamics to bound, from below and from above, the variance of random fluctuations arising from any purely dissipative nonlinear element. The lower bound happens to be a straightforward generalisation of Johnson-Nyquist's formula (fluctuation-dissipation theorem for purely dissipative systems), showing that nonlinearities can only increase the amplitude of noise compared to this Johnson-Nyquist's extended formula. The upper bound departs from the lower bound by a term which involves the skewness of the fluctuations, showing that symmetric fluctuations respect Johnson-Nyquist's nonlinear extension.

We first focus explicitly on the case of  nonlinear electric resistances. The white noise arising due to thermal motion of free electrons in usual linear resistances is always thought of as Gaussian, with intensity given by Johnson-Nyquist's theorem \cite{Johnson1928,Nyquist1928,Davenport1958}.
On the other hand, the noise appearing in some specific devices such as pn junction diodes or tunnel junctions is generally assumed to be a Poisson noise (or shot noise), as confirmed empirically \cite{Reulet2010}.
The exact nature of the white noise in MOS transistors is much more subject to debate, as we shall see.  In this work, provided with results from stochastic thermodynamics, we suggest that the white noise in a MOS transistor ---one of the most important element of modern technologies--- cannot be Gaussian, as implicitly assumed by the widely used semiconductor and circuit-simulation theory\cite{Tsividis2011}, and in some cases cannot be Poisson either.

Despite the focus on electric resistances, our results are general, as they apply to any sort of ideal bath --- thermal or not, electric or not. We provide a toy example on chemical reaction modelling, showing the thermodynamic inconsistency of a Chemical Langevin Equation model.

More generally, these early results call in our view for design principle for stochastic modelling of physical or chemical systems that would guarantee the consistency with the rules of (stochastic) thermodynamics, in complement to existing work such as \cite{rajpurohit2017stochastic}. 



The article is structured as follows. First we review the mathematical theory of white noise, then a key principle of stochastic thermodynamics (local detailed balance) and its consequences on the white noise that are physically admissible.  We then obtain bounds on the variance of noise. We illustrate the diode and the MOS transistor, and a simple chemical reaction. Finally we conclude.

\section{White noise processes}

We consider a resistance as a two-terminal system which, subject to a constant voltage (difference of potential) $V$ at its 
terminals, yields a random current $I(t)$, which behaves as a stationary white noise process whose characteristics depend on the value of $V$. The instantaneous power absorbed by the resistance is $VI(t)$. A note on vocabulary: here we consider the mean $g(V) = \mathbb{E}I$ as part of the white noise process, because mathematically speaking it is cumbersome to artificially separate it from higher moments. This is consistent with the mathematical theory of (Lévy) white noise processes, as detailed e.g. in \cite{barndorff2001levy}. Other references may separate  $g(V) = \mathbb{E}I$ as the `dc component' or `response', reserving the name of `noise' to the zero-mean time-fluctuating signal $I(t)-g(V)$.

Let us call the $\Delta q=\int_0^{\Delta t} I(t) dt$ the net charge (or current increment) traversing the resistance over a time interval $\Delta t$. That $I(t)$ is a white noise process essentially means that the increments over nonoverlapping time intervals are all independent. Recall that the moment generating functions of the sum of two independent random variables $X$ and $Y$ is the product of individual moment generating functions: $\mathbb{E}e^{(X+Y)s}=\mathbb{E}e^{Xs}\mathbb{E}e^{Ys}$ for all $s$ where these quantities are defined. Along with the stationarity of the process, it results that  $\ln \mathbb{E} e^{s \Delta q}$ is proportional to $\Delta t$.   In other words, for some function $M(s)$ called the \emph{cumulant generating function} of the white noise, we have 
\begin{align}
	\mathbb{E} e^{\Delta q s} = e^{M(s) \Delta t} = 1+ M(s) \Delta t + \mathcal{O}(\Delta t^2).   
\end{align}
Thus $M(s)$ collects the nonvanishing part of moments of increment $\Delta q$ in the limit of small times $\Delta t$. 
A well-known example is the Gaussian white noise, characterised by 
\begin{align}
	M(s)= \mu s + \sigma^2 s^2/2.
\end{align}
Here only the mean $\mu= \mathbb{E}\Delta q/\Delta t$ and variance $\sigma^2=\text{Var}(\Delta q)/\Delta t$ do not vanish in the small time limit, while all  moments of order $\ell \geq 3$ vanish:
\begin{align}\lim_{\Delta t  \to 0} \mathbb{E}\Delta q^\ell/\Delta t =0.\end{align}

Another well-known noise is the Poisson noise, obeying $M(s)= \lambda (e^{q_es}-1)$ where  $\lambda >0$ is the expected rate of arrivals  and $q_e$ is the (positive or negative) elementary increment carried by each arrival. 

For instance, if we consider the current through a resistance as a flow of independent charge-carriers, then $q_e$ is the (positive) charge of the electron. The $\ell$th moment $\mathbb{E}\Delta q^\ell$ scale as $q_e^\ell \lambda \Delta t$ for small $\Delta t$. Thus none of the moments can be neglected in the small time limit, in constrast with the Gaussian case. 

One can compose more examples of white noises by addition of independent white noises. The sum of two noises of cumulant generating functions $M_1(s)$ and $M_2(s)$ has cumulant generating function  $M(s)=M_1(s)+M_2(s)$.

For instance, the current through a resistance can sometimes be modelled as a bidirectional Poisson process, where charge-carriers flow in one direction with a rate $\lambda^+$ (positive charge $q_e$) and in the opposite direction with a rate $\lambda^-$ (charge $-q_e$). Here we have $M(s)= \lambda^+ (e^{q_es}-1)+\lambda^-(e^{-q_es}-1)$. Such a noise is called \emph{shot noise} in the electronic literature. The $\ell$th moment in the limit of small $\Delta t$ is 
\begin{align}\label{eq:shotmoment}
	q_e^\ell(\lambda^+ +  (-1)^\ell \lambda^-)\Delta t + \mathcal{O}(\Delta t^2)
	\text{.}
\end{align}



\section{Local detailed balance and its consequences}

We now introduce the constraints imposed by stochastic thermodynamics on the characteristics of the white noise produced by a constant-temperature (possibly nonlinear) resistance. 

A well-known consequence of microscopic reversibility is that the white noise generated by interaction with a thermal bath obeys, under broad physical circumstances, the so-called local detailed balance relation \cite{maes2021local,van2013stochastic}. In the case of an electric resistance subject to a constant difference of potential $V$, it imposes the following constraint on the probabilities (or probability densities) to observe a charge increment $\Delta q$ along an arbitrary time interval $\Delta t$:
\begin{align}\label{eq:ldb}
	\frac{\text{Prob}[\Delta q=-x]}{\text{Prob}[\Delta q=x]}=e^{-V x/kT}.
\end{align}
Here $k$ is Boltzmann's constant and $T$ is the temperature.
Note that $V \Delta q$ is the energy dissipated into the bath (the \emph{heat}) during time $\Delta t$, and 
$V \Delta q/kT$ is physical entropy generated by this exchange of heat (divided by Boltzmann's constant in order to get a dimensionless quantity).

Averaging \eqref{eq:ldb} over all possible $x$ (i.e. summing each term with weight $\text{Prob}[\Delta q=x]$), we recover the so-called \emph{fluctuation relation}, valid for all times $\Delta t$:
\begin{align}\label{eq:fr}
	\mathbb{E}e^{-V \Delta q/kT}  =1.
\end{align}
Taking the logarithm:
\begin{align}\label{eq:fr2}
	M(-V/kT)=0.
\end{align}




The fluctuation relation \eqref{eq:fr} can be developed in Taylor series and yields, after isolating the first order term:
\begin{align}\label{eq:frTaylor}
	g(V) \Delta t=\mathbb{E}\Delta q &=\mathbb{E}\Delta q^2 (V/kT)/2\\&- \mathbb{E}\Delta q^3(V/kT)^2/6\nonumber\\&+ \mathbb{E}\Delta q^4(V/kT)^3/24 - \ldots\nonumber
\end{align}
This formula (valid for all times $\Delta t$) indicates that the response (mean current) $g(V)$ excited by a constant force $V$ can be expressed in terms of the higher moments of the noise. 




\section{Bounding the Variance of Fluctuations}

In engineering situations, especially in electronics, very often \eqref{eq:frTaylor} cannot be used directly because the operating point of interest is far into the nonlinear domain, and because the mean response is easier to model theoretically and measure empirically than the higher moments of fluctuations around the mean. In practice the variance is of most immediate interest in quantifying these fluctuations. 
The purpose of this section is thus to exploit \eqref{eq:fr} in order to bound the variance of fluctuations from below and from above.

We first observe that for all real $x$, the exponential is above its odd-order approximations, including 
\begin{align}
	e^x \geq 1+x \,\,\,\text{and}\,\,\,e^x \geq 1+x + x^2/2+ x^3/6.
\end{align}
Applying these inequalities to \eqref{eq:fr}, we find respectively:
\begin{equation}
	V\mathbb{E}\Delta q \geq 0
\end{equation}
and
\begin{align}\label{eq:upbound}
	\mathbb{E}\Delta q^2 \leq 2kT g(V) \Delta t / V + \mathbb{E}\Delta q^3 (V/3kT).
\end{align}
The first inequality merely expresses the dissipativity of the resistance: energy is always absorbed on average. 

Regarding a lower bound, we resort to the recent Thermodynamic Uncertainty Relations, which is a family of bounds on the variance of some observables in terms of the average entropy production, which is here $V \mathbb{E} \Delta q/kT$. Applying the main result as reviewed in \cite{horowitz2020thermodynamic} to the observable $\Delta q$, along any $\Delta t$:
\begin{equation}
	\label{eq:tur-dq0}
	\frac{ (\mathbb{E}\Delta q)^2}{\text{Var}\Delta q}\leq \frac{1}{2} \frac{ V \mathbb{E} \Delta q}{kT}
\end{equation}
From \eqref{eq:tur-dq0} we deduce, for any $\Delta t$:
\begin{align}\label{eq:tur-dq}
	\text{Var} \Delta q \geq 2kT \mathbb{E}\Delta q/V
\end{align}

Combining \eqref{eq:upbound} and \eqref{eq:tur-dq},
we obtain the main theoretical result of this paper:
\begin{equation}
	\label{eq:main}
	2 k T \, \Delta t \, \frac{g(V)}{V}
	\leq \mathbb{E}\Delta q^2
	\leq 2 k T \, \Delta t \, \frac{g(V)}{V}  + \mathbb{E}\Delta q^3 \, \frac{V}{3kT}
	\text{.}
\end{equation}
Although it is valid for all $\Delta t$, it is the tightest and easiest to interpret in the low $\Delta t$ limit, where $\mathbb{E}\Delta q^2$ coincide with the variance and $\mathbb{E}\Delta q^3$ with the skewness (centred third moment). 
The lhs member of \eqref{eq:main} is a straightforward generalisation of Johnson-Nyquist's formula over the voltage range $[0,V]$ in lieu of linear conductance $G$ for the linear case $g(V)=GV$ --- we call it Johnson-Nyquist's \emph{extended} formula. 
We see that the third moment controls how much nonlinear fluctuations can depart from  Johnson-Nyquist's extended formula. In particular if fluctuations are symmetric around the mean (as in Gaussian noise), then the third-order term in the rhs of \eqref{eq:main} vanishes and subsequently the extended (i.e. nonlinear) Johnson-Nyquist's formula is the only possible. 

\section{The case of shot noise}\label{sec:shot}

An outstanding example of non-Gaussian noise is bidirectional Poisson noise, also called shot noise in the electronic literature. We assume the current results from two opposite Poisson processes of intensities $\lambda^+(V)$ (contributing positively to the current) and $\lambda^-(V)$ (contributing negatively). From local detailed balance \eqref{eq:ldb}  we find that 
\begin{equation}
	\label{eq:ldb lambdas} 
	\frac{\lambda^-(V)}{\lambda^+(V)}=e^{-q_eV/kT}
\end{equation}
The successive moments, in the limit of small times and up to smaller order in $\Delta t^2$, are thus, from \eqref{eq:shotmoment}:
\begin{align}
	\mathbb{E}\Delta q/\Delta t&=g(V)= q_e (\lambda^+ - \lambda^-)
	\label{eq:E(dq) poisson}\\
	\mathbb{E}\Delta q^2/\Delta t&= q_e^2 (\lambda^+ + \lambda^-) = q_e g(V)\frac{1+e^{-q_eV/kT}}{1-e^{-q_eV/kT}} \label{eq:E(dq2) poisson}\\
	\mathbb{E}\Delta q^3/\Delta t&= q_e^3 (\lambda^+ - \lambda^-)= q_e^2 g(V) 
	\label{eq:E(dq3) poisson}
\end{align}

\begin{figure}
	\centering
	\psfragscanon
	\psfrag{V/phiT}[cc][cc]{$V/(kT/q_e)$}
	\psfrag{E(dq2)}[cc][cc]{$(\mathbb{E}\Delta q^2/\Delta t) / \big(2 k T \, g(V)/V\big)$}
	\psfrag{0}[cc][cc]{$0$}
	\psfrag{2}[cc][cc]{$2$}
	\psfrag{4}[cc][cc]{$4$}
	\psfrag{6}[cc][cc]{$6$}
	\psfrag{8}[cc][cc]{$8$}
	\psfrag{10}[cc][cc]{$10$}
	\psfrag{Noise variance}[tl][tl]{\color{black}
		\setlength{\tabcolsep}{0pt} 
		\begin{tabular}{c}
			Noise variance \\
			$\displaystyle \frac{\mathbb{E}\Delta q^2/\Delta t}{2 k T \, \frac{g(V)}{V}}$
		\end{tabular}
	}
	\psfrag{JN}[tl][tl]{\color{blue} 
		\setlength{\tabcolsep}{0pt} 
		\begin{tabular}{c}
			Lower bound \\
			Johnson-Nyquist
		\end{tabular}
	}
	\psfrag{Upper bound}[tl][tl]{\color{red} 
		\setlength{\tabcolsep}{0pt} 
		\begin{tabular}{c}
			Upper bound \\
			$\displaystyle 1 + \frac{\mathbb{E}\Delta q^3/\Delta t}{2 k T \, \frac{g(V)}{V}}$
		\end{tabular}
	}
	\includegraphics[scale=1]{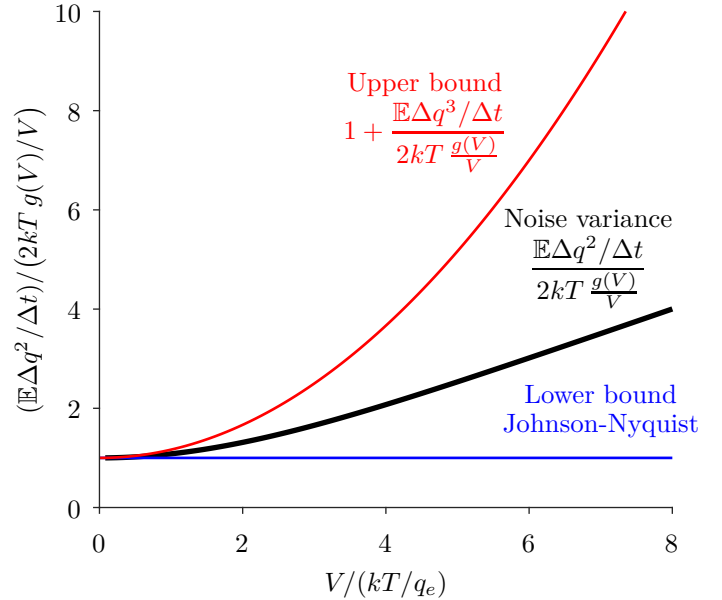}
	\caption{
		Illustration of the main result \eqref{eq:main} applied to the shot noise.
	}
	\label{fig_shot_noise}
\end{figure}

Provided with 
\cref{eq:E(dq) poisson,eq:E(dq2) poisson,eq:E(dq3) poisson},
we illustrate the inequalities \eqref{eq:main} in \cref{fig_shot_noise}. The noise variance (in thick black) has been divided by the lower bound, and this curve shows how the actual noise variance takes off from the extended Johnson-Nyquist noise variance with larger $V$. The upper bound is also plotted (top red curve) in \cref{fig_shot_noise}.

The upper and lower bound in \eqref{eq:main} differ by a factor $1+(q_eV/kT)^2/6$, and this ratio precisely correspond to the top curve in \cref{fig_shot_noise}.
For small $V \ll kT/q_e$ (around 26 mV at room temperature), the Johnson-Nyquist bound is a good approximation of the variance.

An example where the shot noise model is applicable is Shockley's model for the pn-junction diode, where  $\lambda^-$ is considered constant for all $V$. The mean current is thus exponentially increasing for positive $V$, while it saturates to $q_e \lambda^-$ for negative values. 
Thus we have \cite{Wyatt1999}:
\begin{align}
	g(V)=I_0(e^{\frac{Vq_E}{kT}}-1)
\end{align}
where $I_0=q_e\lambda^-$ is the (usually small) negative current obtained for a strong negative voltage. For a strong positive voltage on the other hand the current grows exponentially (in the limit where this model is valid). This strong asymmetry is the  landmark feature of the diode.

\section{The case of MOS transistor}

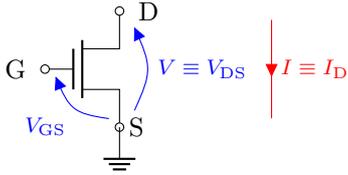
\begin{figure}[!t]
	\centering
	\begin{circuitikz}[european voltages, transform shape, line cap=rect, nodes={line cap=butt},scale=1]
		\newcommand\Icolor{red}
		\newcommand\Vcolor{blue}
		\draw
		(0,0) node[nmos] (nMOS) {}
		(nMOS.D) to[short,-o] ++(0,0) node[label={[font=\normalsize,color=black]right:D}] (nD) {}
		(nMOS.G) to[short,-o] ++(0,0) node[label={[font=\normalsize,color=black]left:G}] (nG) {}
		(nD)++(0.2,0) node[] (nDD) {}
		(nD)++(2,0) node[] (nDDD) {}
		(nMOS.S) to[short,-o] ++(0,0) node[ground,scale=1,label={[font=\normalsize,color=black]right:S}] (nS) {}
		(nS)++(2,0) node[] (nSS) {};
		{\color{\Icolor}
			\ctikzset{current/distance = 0.5}
			\draw
			(nDDD) to[short, font=\small, i=\textcolor{\Icolor}{$I \equiv \ID$}, color=\Icolor] (nSS);
		}
		{\color{\Vcolor}
			\ctikzset{voltage/bump b=0.5}
			\draw
			(nS) to[open, font=\small, v^=\textcolor{\Vcolor}{$\VGS$}, color=\Vcolor] (nMOS.G)
			(nS)+(0.2,0) to[open, font=\small, v=\textcolor{\Vcolor}{$V \equiv \VDS$}, color=\Vcolor] (nDD)
			;
		}
	\end{circuitikz}
	\caption{MOS transistor schematic: notations for the voltages and the current.}
	\label{fig_MOS}
\end{figure}


The MOS (Metal-Oxide Semiconductor) (field-effect) transistor is the key 
element of integrated circuits in silicon.
As depicted in \cref{fig_MOS}, a MOS transistor can be considered as a \emph{three}-access structure.
The device can be seen as a \emph{voltage-controlled} nonlinear resistance between two longitudinal accesses, the source (S) and the drain (D).
The positive current flows from the drain to the source (by convention). 
We denote it by $I \equiv \ID$ and we refer to it as the \emph{drain(-to-source) current}, while the longitudinal voltage difference is $V \equiv \VDS$.
The gate G terminal acts to control the current and the conductance; $\VGS$ is always implicitly assumed to be fixed thereafter, thus acts as a mere parameter on the nonlinear source-drain resistance.


Although the device characteristics are necessarily continuous, it is convenient to study two important regions of operation separately: the \emph{weak} and \emph{strong inversion}, respectively. Conceptually, a suitably defined \emph{threshold voltage} $\Vth$ abruptly marks the separation between the two regions.

The weak-inversion region ($\VGS \leq \Vth$) is also called \emph{subthreshold} region. In this region, models and experiments confirm that:
\begin{align}
	g(V)=I_0  (1-\exp(-Vq_e/kT))\\
	\mathbb{E}\Delta q^2/\Delta t=I_0  (1+\exp(-Vq_e/kT))
\end{align}
for some $I_0$ depending on the parameter $\VGS$ and the characteristics of the transistor. This is compatible with a shot noise with a constant $\lambda^+$ (depending on fixed $\VGS$ but not $V$)\cite{Sarpeshkar1993,Tsividis2011, Wyatt1999,Tedja1994,freitas2021stochastic}, similarly to the pn junction case mentioned above, in accordance with \cref{eq:ldb lambdas,eq:E(dq) poisson,eq:E(dq2) poisson}. Another standard derivation proceeds from an implicit Gaussian assumption \cite{Tsividis2011}, with the same result for first and second moment. Nevertheless, as we have seen in the previous section, a Gaussian assumption is not thermodynamically correct here, as it automatically implies Johnson-Nyquist's extended formula.

The strong-inversion roughly corresponds to $\VGS \geq \Vth$.
The classical `long-channel' MOS transistor theory predicts the following model for average drain current \cite{Tsividis2011}:
\begin{equation}
	\label{eq:MOS g(V)}
	g(V) = 
	\begin{cases}
		\displaystyle \beta \, \Big((\VGS - \Vth)V-\frac{V^2}{2}\Big) \\ \quad\text{if $V < \Vsat=(\VGS-\Vth)$} \\
		\displaystyle \beta \, \frac{\Vsat^2}{2}
		\quad\text{if $V \geq \Vsat$ (saturation regime).}
	\end{cases}
\end{equation}
$\beta$ is a constant which absorbs 
some transistor and physical parameters.
The variance of the white noise in these regions is  usually written as \cite{Tsividis2011, Tedja1994}:
\begin{equation}
	\label{eq:MOS noise strong inv}
	\mathbb{E}\Delta q^2/\Delta t
	= 2 kT \, \beta (\VGS - \Vth) \, \frac{2}{3} \frac{1+\eta+\eta^2}{1+\eta}
\end{equation}
where 
\begin{equation}
	\label{eq:eta}
	\eta = 
	\begin{cases}
		\displaystyle 1 - \frac{V}{\Vsat} \quad \text{if } V \leq \Vsat \\
		\displaystyle 
		\mathrlap{0} \hphantom{1 - \frac{V}{\Vsat}} \quad \text{if } V \geq \Vsat
		\text{.}
	\end{cases}
\end{equation}

Particularly, in saturation, we have $\eta = 0$ and \eqref{eq:MOS noise strong inv} reduces to
\begin{equation}
	\label{eq:MOS noise strong inv sat}
	\mathbb{E}\Delta q^2/\Delta t
	= 2 kT \, \frac{2}{3} \, \beta (\VGS - \Vth) =  2 kT \, \frac{2}{3} \, \beta \Vsat
	\text{.}
\end{equation}
Additional non-equilibrium effects, for instance the charge carrier velocity saturation and carrier heating in the channel, can inflate the noise \cite{Tsividis2011,Han2004}. The prediction  given by \eqref{eq:MOS noise strong inv sat} may therefore be regarded as 
a fair lower bound for the noise variance in saturation.

As a sanity check, we consider the small $V \ll V_\text{sat}$ limit. We have $\eta \approx 1$ and \eqref{eq:MOS noise strong inv} yields
\begin{equation}
	\label{eq:MOS noise strong inv V = 0}
	\mathbb{E}\Delta q^2/\Delta t
	= 2 kT \, \beta (\VGS - \Vth)  
	\text{.}
\end{equation}
Neglecting the quadratic $V^2/2$ term in \eqref{eq:MOS g(V)}, we identify $\beta (\VGS - \Vth)$ with the average conductance $g(V)/V$ and we find as expected:
\begin{equation}
	\label{eq:MOS noise strong inv V = 0 cond}
	\mathbb{E}\Delta q^2/\Delta t
	= 2 kT \, \frac{g(V)}{V} 
	\text{,}
\end{equation}
which is exactly the 
Johnson-Nyquist's formula, the lower bound of \eqref{eq:main}, as expected.



\begin{figure}
	\centering
	\psfragscanon
	\psfrag{V/Vsat}[cc][cc]{$V/\Vsat$}
	\psfrag{Sid/(2 kT beta (VGS-Vth))}[cc][cc]{$(\mathbb{E}\Delta q^2/\Delta t) / \big(2 kT \, \beta (\VGS-\Vth)\big)$}
	\psfrag{0}[cc][cc]{$0$}
	\psfrag{1/2}[cc][cc]{$1/2$}
	\psfrag{2/3}[cc][cc]{$2/3$}
	\psfrag{1}[cc][cc]{$1$}
	\psfrag{Noise model}[tl][tl]{\color{black} Noise model \eqref{eq:MOS noise strong inv sat}}
	\psfrag{Johnson-Nyquist}[tl][tl]{\color{blue} Johnson-Nyquist}
	\includegraphics[scale=1]{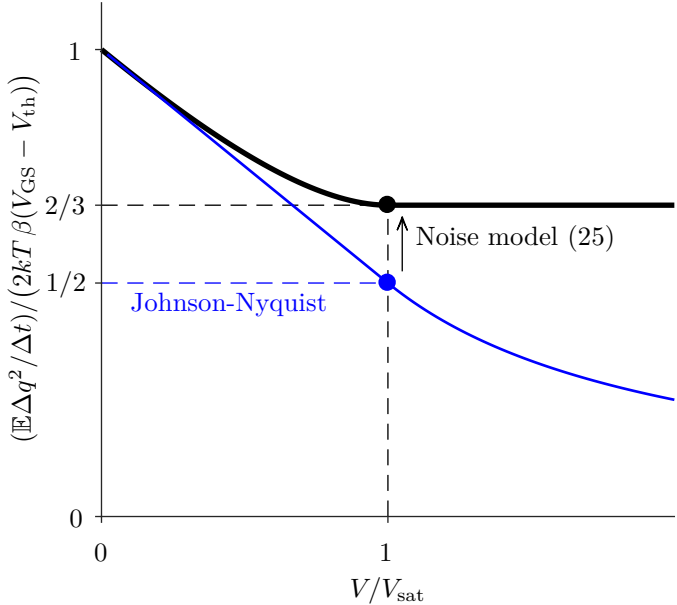}
	\caption{
		MOS transistor white noise variance in strong inversion, compared to the extended Johnson-Nyquist formula.
	}
	\label{fig_MOS_Sid}
\end{figure}

As $V$ grows, the mismatch between the noise variance \eqref{eq:MOS noise strong inv sat} and the extended Johnson-Nyquist lower bound  (illustrated in \cref{fig_MOS_Sid})  grows as well. This proves, using \eqref{eq:main}, that fluctuations have nonzero skewness, thus are asymmetric. In particular they are not Gaussian, unlike implicitly assumed in the standard derivation \cite{Tsividis2011} of \eqref{eq:MOS noise strong inv}.

Since \eqref{eq:MOS noise strong inv} does not satisfy the relation \eqref{eq:E(dq2) poisson}, it is not a shot noise either. Thus the white noise in the MOS transistor in strong inversion regime offers an example of white noise in electronics that is neither Gaussian nor shot noise, with a non-negligible impact of the third moment. 

It is instructive to compute the order of magnitude of the third moment. We limit this numerical example to the strong-inversion region and the saturation regime, for the sake of simplicity. 
In this case the extended Johnson-Nyquist's term is, from \eqref{eq:MOS g(V)}:
\begin{equation}
	\label{eq:JN saturation}
	2 k T \, \frac{g(V)}{V} = 2 k T \, \beta \Vsat \, \frac{1}{2}\frac{\Vsat}{V}
	\text{.}
\end{equation}
Substituting \eqref{eq:MOS noise strong inv sat} and \eqref{eq:JN saturation} in \eqref{eq:main} leaves us with
\begin{equation}
	\label{eq:MOS third moment}
	\frac{\mathbb{E}\Delta q^3}{\mathbb{E}\Delta q^2} \geq \frac{3kT}{V} \, \big( 1 - \frac{3}{4} \frac{\Vsat}{V} \big)
	\text{.}
\end{equation}
\Cref{eq:MOS third moment} provides a lower bound for the third-order moment (i.e. the skewness) of current noise, relatively to the variance. We choose this ratio because it takes a simple form. At  $V = \frac{3}{2}\Vsat$, the rhs reaches  a maximum:
\begin{equation}
	\label{eq:MOS third moment max}
	\frac{\mathbb{E}\Delta q^3}{\mathbb{E}\Delta q^2} = \frac{kT}{\Vsat} = q_e \frac{kT/q_e}{\Vsat}
	\text{.}
\end{equation}
In \eqref{eq:MOS third moment max}, we have expressed the ratio between skewness and variance as a number of elementary charges. At room temperature $kT/q_e \approx \SI{26}{\milli\volt}$.
Regarding $\Vsat$, it typically ranges from a few $\si{\volt}$ in old $\si{\micro\meter}$ CMOS technologies and down to several hundreds of $\si{\milli\volt}$ in the most advanced decananometer technologies.
$\Vsat = \SI{1}{\volt}$ is thus a realistic value for our example. We finally find $\sim q_e / 40$ as a lower bound for the skewness to variance ratio. 

To appreciate this result, we observe from \eqref{eq:E(dq2) poisson} and \eqref{eq:E(dq3) poisson} that the absolute value of skewness-to-variance ratio for any shot noise ranges from $0$ (for $V \ll kT/q_e$) to $q_e$ (for $V \gg kT/q_e$).  This has been experimentally confirmed for a tunnel junction \cite{Reulet2010}.


\section{Beyond Electronics: the Inadequacy of Gaussian Noise for Chemical Reaction Systems}
Although we adopted the vocabulary and examples in the field of electronics, the main formula \eqref{eq:main} has a validity much beyond fluctuations appearing in electric resistances and non electronic devices. It applies, up to adaptation of vocabulary, to any dissipative process into a 
thermodynamic reservoir (a heat bath, a particle bath, etc.). For example it applies to dissipation of mechanical energy to a bath of temperature $T$ through friction, or a diffusion of Brownian particle in a fluid  (the historical context of introduction of fluctuation-dissipation relation by Einstein). Here we provide a supplementary example borrowed to reactional systems such as used in chemistry, where we emphasise the analogy with the diode studied earlier in this paper.

Consider the following chemical reaction, describing the creation or destruction of molecular species $A$ out of an unspecified substrate (or particle reservoir):
\begin{equation}\label{eq:0A}
	\emptyset \rightleftharpoons A.
\end{equation}
The stochastic dynamics is characterized by a certain concentration (number of molecules) $x_A$. We assume creation of molecules of $A$ occurs according to a Poisson process $\lambda^+$, and destruction of molecules of $A$ with a rate $\lambda^-$. 

According to the law of mass action ---a common model of kinetics for chemical reactions--- we expect that $\lambda^+$ is constant regardless of $x_A$, while $\lambda^-$ is proportional to $x_A$ (which amounts to assuming that each molecule of $A$ may degrade with some constant probability independently of other molecules), although we make little use of this fact here.

We assume that the concentration $x_A$ is maintained constant in a way that is not modelled here, as the molecules of $A$ are continuously evacuated or restored in order to compensate the natural dynamics of \eqref{eq:0A}. In the chemical language we say that $A$ is \emph{chemostated}.

This example brings us to state a more general version of the local detailed balance condition \eqref{eq:ldb}. The flow of mass $\Delta x_A$ replaces here the flow of charge carriers. We now write the local detailed balance (over any interval $\Delta t$) as:
\begin{align}\label{eq:ldb2}
	\frac{\text{Prob}[\Delta x_A=-x]}{\text{Prob}[\Delta x_A=x]}=e^{-\Delta \sigma}
\end{align}
where $\Delta \sigma$ is the entropy production associated to the flow $\Delta x_A$. This entropy production, which amounts to $V\Delta q/kT$ in case of the electrical resistance, is  written analogously for a reaction like \eqref{eq:0A}  in the form
\begin{align}
	\Delta \sigma=\mu_A \Delta x_A/kT,
\end{align}
defining the \emph{chemical potential} $\mu_A$---a function of $x_A$ in general, thus a constant in our case study. In both cases, a \emph{thermodynamic force} ($V/kT$ or $\mu_A/kT$) generates a \emph{flow} ($\Delta q$ or $\Delta x_A$), associated with an entropy production $\Delta \sigma$, product of the force by the flow. See \cite{rao2016nonequilibrium} for a stochastic thermodynamic treatment of general chemical reactional systems.

We want to the characterise the flow $\Delta x_A$ generated by a constant chemical potential $\mu_A$ (associated to the constant concentration $x_A$). In this context, the main formula \eqref{eq:main} thus becomes
\begin{equation}
	\label{eq:mainchem}
	2 k T \,  \frac{\mathbb{E}\Delta x_A}{\mu_A}
	\leq \mathbb{E}\Delta x_A^2
	\leq 2 k T \,  \frac{\mathbb{E}\Delta x_A}{\mu_A}  + \mathbb{E}\Delta x_A^3 \, \frac{\mu_A}{3kT}
	\text{.}
\end{equation}
It is commonly accepted that a microscopically `correct' model noise under broad conditions is a bidirectional Poisson noise, where one molecule of $A$ is created or destroyed, leading to a countable-state Markov chain description for the evolution of $x_A$, called the Chemical Master Equation \cite{gillespie1992rigorous}. Under the law of mass action evoked above, the rate of creation $\lambda_+$ is constant independently of $\mu_A$ thus the form of $\mathbb{E}\Delta x_A^k$ is formally identical to the moments $\mathbb{E}\Delta q^k$ of Shockley's diode or subtreshold MOS transistor, up to the sign.

However, it is customary to replace the bidirectional Poisson model with a Gaussian white noise, so as to use it as a building block to describe full reaction networks with a continuous-state Chemical Langevin Equation to describe the evolution of concentrations of species. 
From the discussion above, we see that this Gaussian assumption, however convenient, is not thermodynamically consistent since the actual noise variance does not follow the nonlinear Johnson-Nyquist's formula far from equilibrium, as argued in Section \ref{sec:shot}. Interestingly, the inconsistency of Gaussian noise assumption (and Chemical Langevin Equation) far from equilibrium has been noticed in \cite{horowitz2015diffusion}, through analytical and numerical computations. We here explain the inconsistency straight away from the main theoretical result \eqref{eq:mainchem} which has been deduced from fundamental principles.
\section{Conclusion}
The traditional assumption that the white noise is Gaussian, notably widely used in the electronic literature for circuit simulation, has here been questioned at the light of recent theoretical results from stochastic thermodynamics.
\Cref{eq:main} shows that, when the noise variance exceeds the extended Johnson-Nyquist prediction for a nonlinear device, then the noise process is necessarily \emph{non} Gaussian. 
The shot noise, i.e. a bidirectional Poisson process, is an important example.
Although circuit-simulation tools predominantly rely on Gaussian noise assumptions, very recent work assessing the impact of white noise in low-voltage MOS memories in the weak inversion regime made use of a Poisson model of thermal fluctuations \cite{Rezaei2020}.
However, to our best knowledge, no experimental work measuring the third-order moment of the white noise in MOS transistors has been published so far, while our theoretical calculation \eqref{eq:MOS third moment} demonstrates its existence.
Neither has the effect of the noise skewness on the circuit behaviour been investigated yet, unlike in chemistry, where it is shown to have an impact on the long term distribution of concentrations \cite{horowitz2015diffusion}.

Beyond their immediate interest, we see this result as a motivation to build a theory of feedback systems that is able to model systems at a mesoscopic scale ---where noise is non-negligible---in full compatibility with thermodynamics.

\bibliography{main}


\end{document}